\documentclass[11pt]{article} 
\usepackage[ansinew]{inputenc} % Acepta caracteres en castellano
\usepackage{amsmath}
\usepackage{amsfonts}
\usepackage{amssymb}

\usepackage[numbers]{natbib}
\usepackage{rotating}
\usepackage{graphicx}

\begin{document}

% Title portion
\title{Astroparticle Physics at Eastern Colombia}

\author{
\textbf{Hern\'an Asorey} \\
\textit{Laboratorio Detecci\'on de Part\'{\i}culas y Radiaci\'on,} \\
\textit{Centro At\'omico Bariloche e Instituto Balseiro (CNEA/UNCuyo)}\\
\textit{San Carlos de Bariloche, R\'io Negro-Argentina and }\\
\textit{Escuela de F\'{\i}sica - Universidad Industrial de Santander} \\ 
\textit{Bucaramanga, Santander-Colombia.} \\
\textbf{ Luis A. N\'u\~nez }\\
\textit{Escuela de F\'{\i}sica - Universidad Industrial de Santander }\\ 
\textit{Bucaramanga, Santander-Colombia and} \\
\textit{Departamento de F\'{\i}sica, Universidad de Los Andes, M\'erida-Venezuela.}
}
 
\maketitle

\begin{abstract}
We present the emerging panorama of Astroparticle Physics at Eastern
Colombia, and describe several ongoing projects, most of them related to
the Latin American Giant Observatory (LAGO) Project. This research work is
carried out at the Grupo de Investigaciones en Relatividad y Gravitaci\'on of
Universidad Industrial de Santander.  
\end{abstract}

% Head 1
\section{INTRODUCTION}

Astroparticle Physics is now one of the most exciting interdisciplinary fields
in High Energy Physics, where particle Physics and Astrophysics share satellites and ground based detectors to explore new phenomena and innovative applications. 

In Colombia, since 2010, at the Universidad Industrial de Santander (UIS, Bucaramanga), the
Relativity and Gravitation Research Group (GIRG for its Spanish acronym for
Grupo de Investigación en Relatividad y Gravitaci\'on) develops a
research line in close contact with researchers of the
Latin American Giant Observatory (LAGO, formerly known as Large Aperture GRB
Observatory) project.

\begin{figure}
\begin{center}
\includegraphics[width=0.60\textwidth]{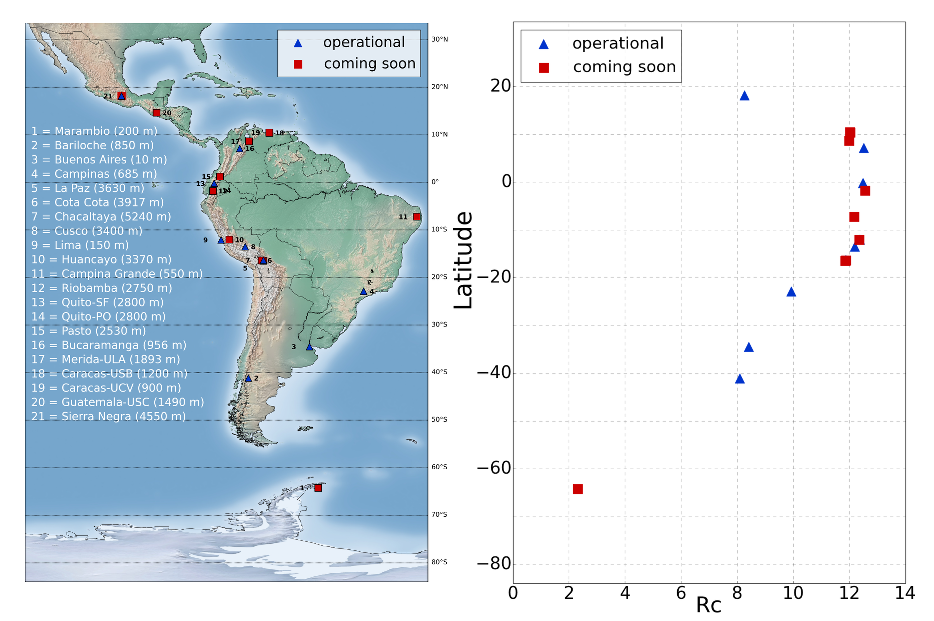} 
\caption{The Sites of the Latin American Giant Observatory (LAGO), currently
located in eight countries at Latin America. The first detectors are in
operation (blue triangles), and some other detectors (red squares), are planned
to start in 2015-2016 \citep{Asorey2015, Sidelnik2015}. In the right panel, the
vertical rigidity cut-off of each site, calculated according the procedure
described in section  \ref{SpaceWeatherProg} and \citep{AsoreyEtal2015C}, are
shown.}
\label{LAGOSitesChacaltaya}
\end{center}
\end{figure}

The Latin American Giant Observatory \citep{Asorey2015} is a highly integrated
and collaborative research and academic project with more than eighty Ibero
American astroparticle researchers, motivated by the experience of the Pierre
Auger Observatory \citep{Auger2015} in Argentina, and devoted to study space
weather effects \citep{AsoreyEtal2015B} and high energy transient phenomena
such as Gamma Ray Bursts (GRB) on ground-based detectors
\citep{AsoreyEtal2015C}. Long-term modulation and transient events can also be
characterized by using the LAGO detection network, as it spans over a big area
with different sites at different latitudes, longitudes and geomagnetic
rigidity cut-offs. Presently LAGO has 10 country members (Argentina, Bolivia,
Brazil, Colombia, Ecuador, Guatemala, Mexico, Peru, Spain and Venezuela)
spanning a non-centralized and collaborative network of 25 institutions with 10
ground-based Water Cherenkov Detector (WCDs) \citep{Sidelnik2015}, located at
different altitudes from Mexico through Patagonia. Other detectors are expected
to be up and running in the near future including two WCD in the Antarctica
Peninsula\citep{DassoEtal2015}, see Figure \ref{LAGOSitesChacaltaya}.

This work is organised as follows: in the next section we describe two new
analysis of the scalers rates (an implementation of the single particle
technique \citep{Vernetto2000}) of the LAGO WCDs installed at Mount Chacaltaya.
One of these analysis is conducted to search short transient events ($\Delta t
\lesssim 1$\,minute) and the other is aimed to find periodic long term signals.
Next, we present the LAGO Space Weather program and show how flux variations of
secondary particles at ground level can be related to heliospheric and
geomagnetic modulation of Galactic Cosmic Rays and provide precise information
of the variable conditions of the near-Earth space environment. With this
toolkit, developed to study the Space Weather phenomena, we are able to easily
determine the total integrated flux of cosmic radiation in any place at the
Earth surface or even, e.g., along specifics commercial flight trajectories.
For this particular case, we calculate the integrated exposure due to the
radiation background during a flight and its modulation by effects such as
altitude, latitude, exposure time and transient space weather events. Following
the description of our projects, we prove the capabilities of single WCD to
detect transient space weather phenomena from ground level. The introduction of
the Multi-Spectral Analysis Technique also shows that using even a single
detector it is possible to determine the flux at bands that are dominated by
different types of particles coming from different primaries contributions.

\section{SEARCHING FOR SHORT/LONG TERM SIGNALS IN LAGO}

In this section we present a selection of our observations which confirm that,
when using an adapted analysis technique to the characteristics of our small
detectors, it is possible to observe, from the ground level and in the LAGO
network of WCDs, different type of signals of different astrophysics nature at
different time scales.

\subsection{Short Term Signals: Searching for Gamma Ray Burst}\label{ShortSignals}

LAGO site in Bolivia is located at Mount Chacaltaya ($16^{\circ}21'00''$S,
$68^{\circ}07'53''$W), at $5270$\,m above sea level, where the instrumented
area and the high altitude of the site provide enough sensitivity to detect
high energy Gamma Ray Bursts (GRBs) \citep{AllardEtal2008}.  Recent studies
\citep{NunezCastineyra2015}, based on CORSIKA simulations \citep{HeckEtal1998},
show that the angular aperture of the Chacaltaya site can be extended up to a
zenith angle of $25^\mathrm{o}$ in the energy range of interest for GRB.
Combining these result with the uptime of each of the three WCD at Chacaltaya
in this period, the total exposure of this site accumulated during the
2010-2012 period was $2.7 \times 10^8$\,m$^2$\,s\,sr \citep{AsoreyEtal2015C}.

Several validations of data quality cuts were introduced in all datasets
collected \citep{NunezQuinonezSarmiento2013} and, by using the moving window
average (MWA) method \citep{Sarmiento2012},  we look for $\geq |3 \sigma|$
deviations in the central $5$\,ms bin on a $2$\,minutes (24,000 time bins). If
such a deviation is observed in at least two operating detectors and is present
in both the low and the intermediate deposited energy sub-channels, the signal
is tagged as a potential transient (GRB) candidate. 

Applying this approach over 2 terabytes of data collected in the period
2010-2012 at Chacaltaya, we found a potential candidate, started on Wed Dic 07
15:45:49.675$\pm$0.005 UTC 2011 (unix time $(1323272749.675 \pm 0.005)$\,s). At
this time, the equatorial coordinates of the zenith at Chacaltaya were RA/Dec
(J2000) $16^\mathrm{h} 17^\mathrm{m} 31.3^\mathrm{s} / -16^\circ 21' 00"$, with
an acceptance aperture of $\theta \lesssim 25^\circ$.  After a careful
examination of the signals shapes, calibration data and operation metadata of
our detectors and the atmospheric database of this site, we discarded the
possibility that this event was produced by detectors malfunctions, HF noises,
electric lightnings or other phenomena of atmospheric origin. 

\begin{figure}[!ht]
\begin{tabular}{cc}
\includegraphics[width=0.48\textwidth]{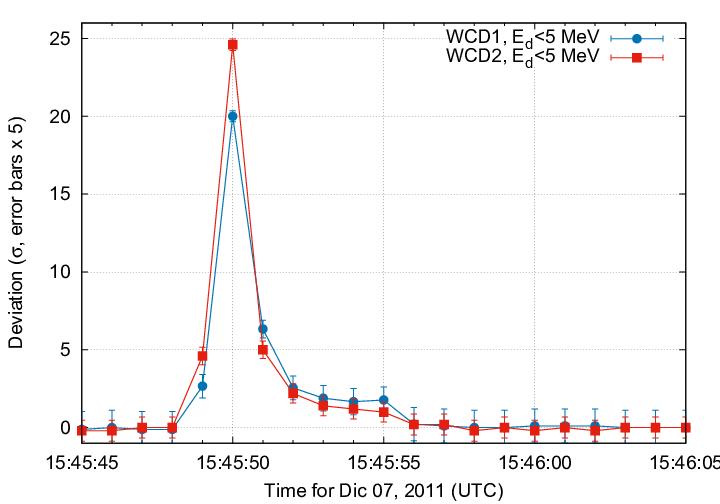} &
\includegraphics[width=0.48\textwidth]{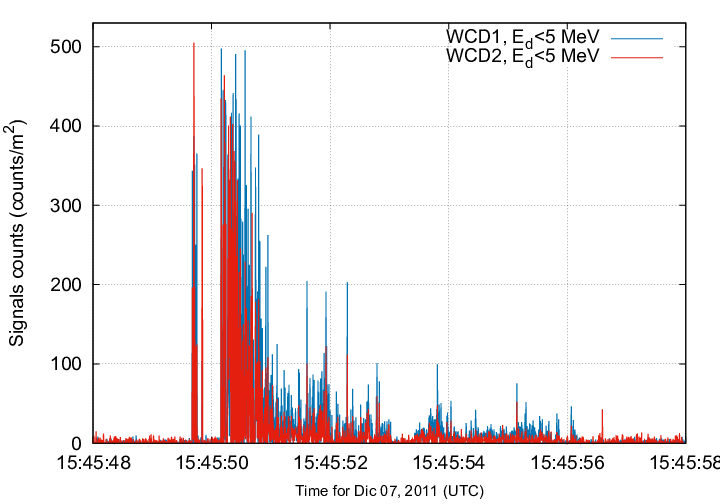} \\
(a) & (b)
\end{tabular}
\caption{Potential candidate signals for the event registered on Wed Dic 07
15:47:02.378 UTC 2011. In panel (a) we show our alert as described in the text,
while in the panel (b) the registered signals in $5$\,ms temporal bins are
shown}
\label{LAGOGRBSignals}
\end{figure}

The detected signals are shown in figures \ref{LAGOGRBSignals}. We did not find
any similar events registered on both the SWIFT \citep{BarthelmyEtal2005} and
Fermi satellites \citep{AceroEtal2015}  and also at the Gamma-ray Coordinates
Network (GCN) database\footnote{{http://gcn.gsfc.nasa.gov/}}. However, at this
time the field of view of Fermi was well outside of our acceptance cone in
Chacaltaya \citep{AsoreyEtal2015C}.

\subsection{Lont Term Signals: Epoch Analysis}
\label{LongSignals}
On the search for periodic signals we performed a series of analyses over large
periods of time on the data collected at the Chacaltaya station. Data stacking
or summation in two different time systems, solar and sidereal, were made over
the data. The idea behind this process is based on the random, poissonian,
nature of the majority of the radiation measured by our detectors. If any not
random, periodic, signal were to exist, and if sufficient data were to be
summed in a scale in which this signal happened at the same time every day or
month or week, an indication of the underlying signal would appear. No matter
how small this signal could be it will build an observable rise on the line
that represent the summed data, as the random fluctuations will eventually
cancel and the signal will grow \cite{NunezCastineyra2015}. This summation of
data were made on two different systems: sidereal time, for signals coming from
outside of our Solar System and in solar time, for signals modulated by Solar
activity. A one-minute average signals were summed with the average of the
corresponding minute of the next day, solar or sidereal.

\begin{figure}[!ht]
\begin{tabular}{cc}
	\includegraphics[width=0.48\textwidth]{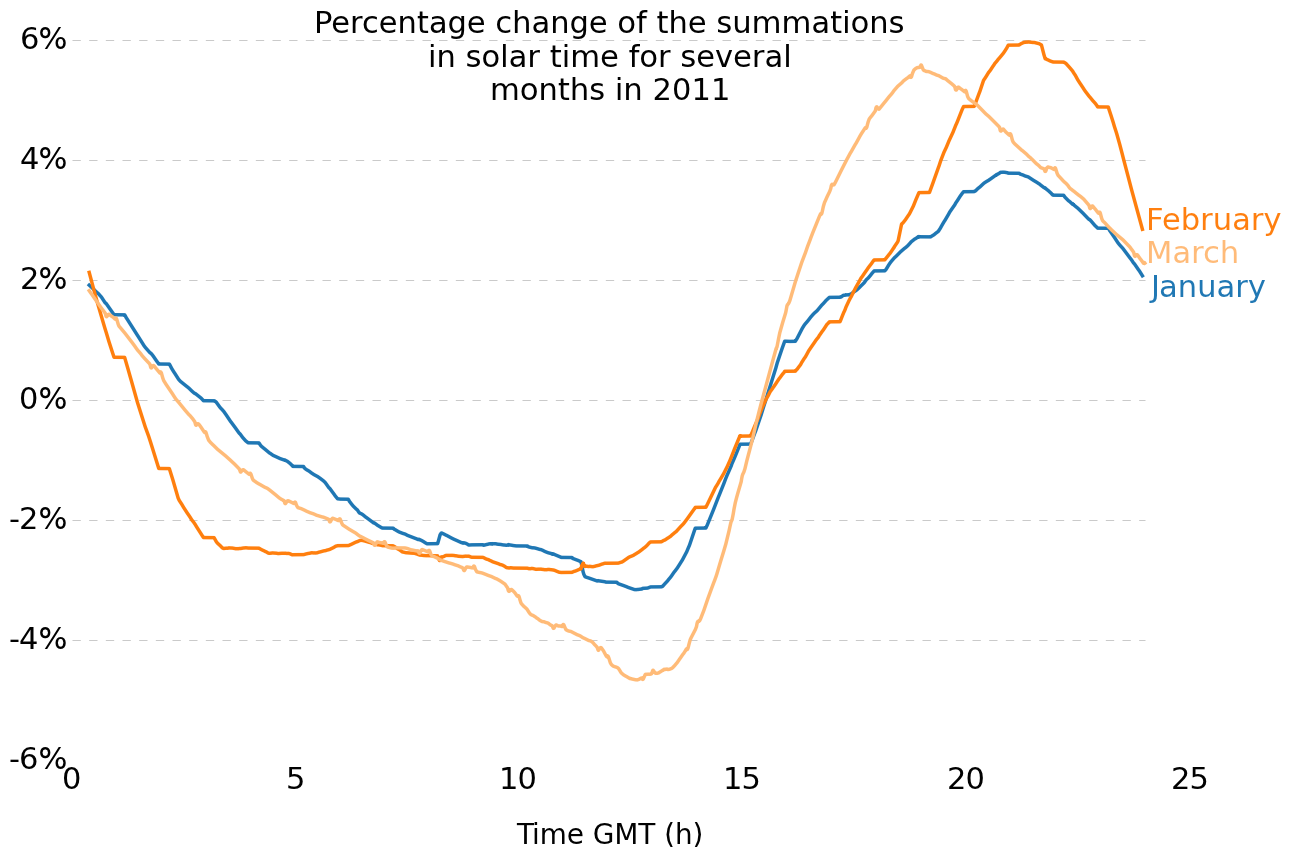} & 
	\includegraphics[width=0.48\textwidth]{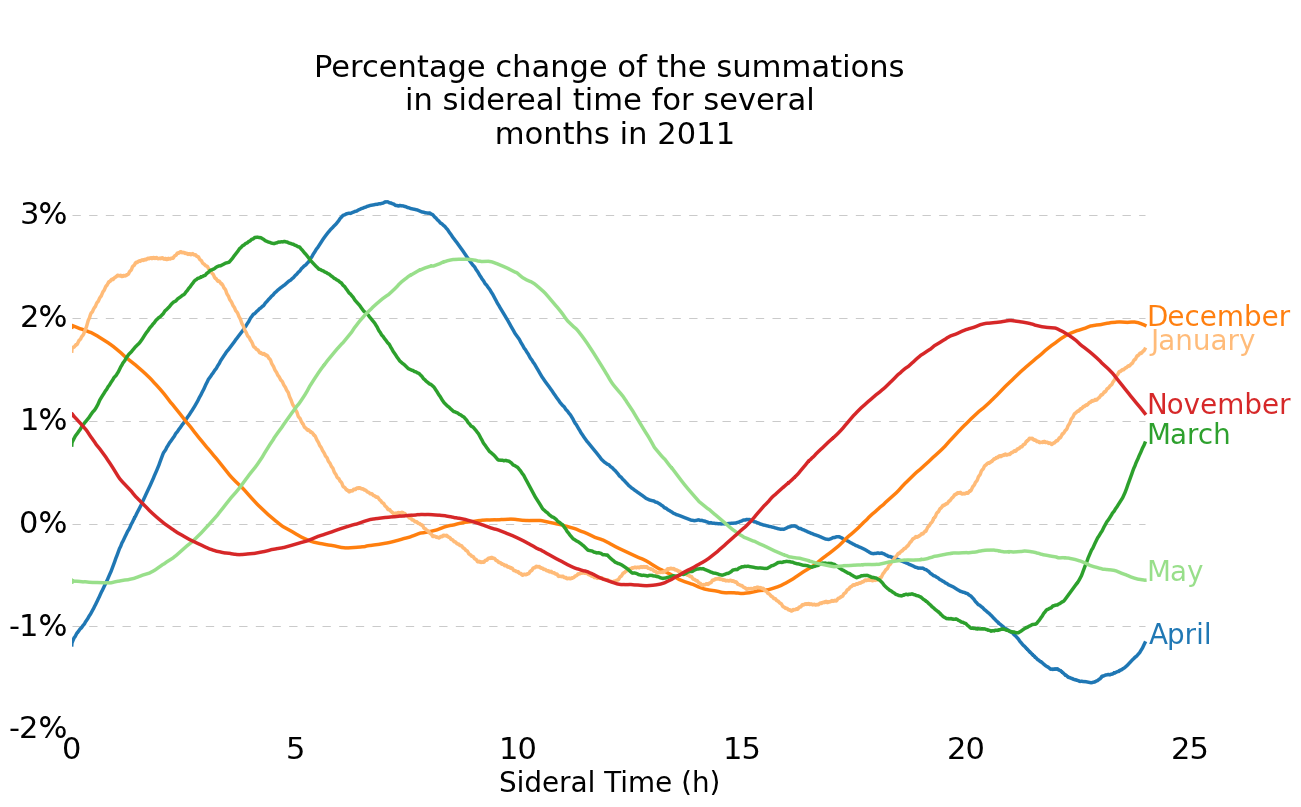} \\
	(a) & (b)
\end{tabular}
\caption{Solar daily modulation in the flux of cosmic rays observed at the LAGO
site in Chacaltaya. This phenomena was observed by stacking and summing the
measured and corrected fluxes in solar (a) and sidereal (b) times.}
\label{SumFig}
\end{figure}

The results of the complete summation on the data collected in the Chacaltaya
station in 2011 is show in Fig. \ref{SumFig} where both solar time and sidereal
times are shown here, and have consistent results. In solar time it appears a rise
on the counts that holds in the same stage of the solar day while in sidereal
time the rise moves about two hour every month. These facts are clear
indicators of the solar nature of this phenomenon. The amplitude and phase
observed on the solar time variations are consistent with the well known daily
modulation of solar origin in the flux of low energy cosmic rays. This result
also shown that our single WCD, combined with the appropriated data acquisition
and analysis techniques, are capable to detect space weather related phenomena
from the Earth surface.

\section{THE LAGO SPACE WEATHER PROGRAM}

\label{SpaceWeatherProg}

The LAGO Space Weather (LAGO-SW) program studies how flux variations of
secondary particles at ground level can be related to heliospheric modulation
of GCRs and provides precise information of variable conditions of the
near-Earth space environment \citep{Asorey2013,AsoreyEtal2015B}. It is
supported by an intensive and detailed chain of simulations, accounting three
important factors: the geomagnetic effects, the development of the extensive
air showers (EAS) in the atmosphere, and the detector response to the different
types of secondary particles at ground level by means of a GEANT4 model
\citep{CalderonAsoreyNunez2015, OtinianoEtAl2015, AsoreyEtal2015B}.

\subsection{Geomagnetic Field Effects}

The geomagnetic field (GF) effects on the propagation of charged particles that
could contribute to the background radiation at ground level can be
characterized by the directional rigidity cut-off $R_c$ at each LAGO site (see
figure \ref{LAGOSitesChacaltaya}), considered as a function of the geographical
position at some altitude (Lat, Lon, Alt) and the arrival direction ($\theta,
\phi$) of the primary at this point. Additionally, a time stamp (TS
YYMMDDHHMMSS UTC) should also be included to address transient phenomena during
those periods of intense geomagnetic activity, so $R_c \equiv R_c
\left(\mathrm{TS},\mathrm{Lat},\mathrm{Lon},\mathrm{Alt},\theta,\phi \right).$

This directional $R_c$ can be calculated at each LAGO site using the
Magnetocosmics code \citep{Desorgher2003} by applying the backtracking
technique \citep{MasiasDasso2014}. We use the International Geomagnetic Field
Reference (IGRF) version 11 \citep{IGRF11} for modeling the near-earth GF ($r <
5 R_{\oplus}$)  and the Tsyganenkov (TSY01) \citep{Tsyganenko2002} to describe
the outer GF. As it is commonly used, our description of the GF can be tagged
by six parameters: solar wind dynamic pressure, D$_{\mathrm{ST}}$ index, $B_y$
and $B_z$ components of the geomagnetic field (in GSM system), and the $G_1$
and $G_2$ parameters of TSY01 model.

\begin{figure}[htb!]
\begin{tabular}{cc}
\includegraphics[width=0.42\textwidth]{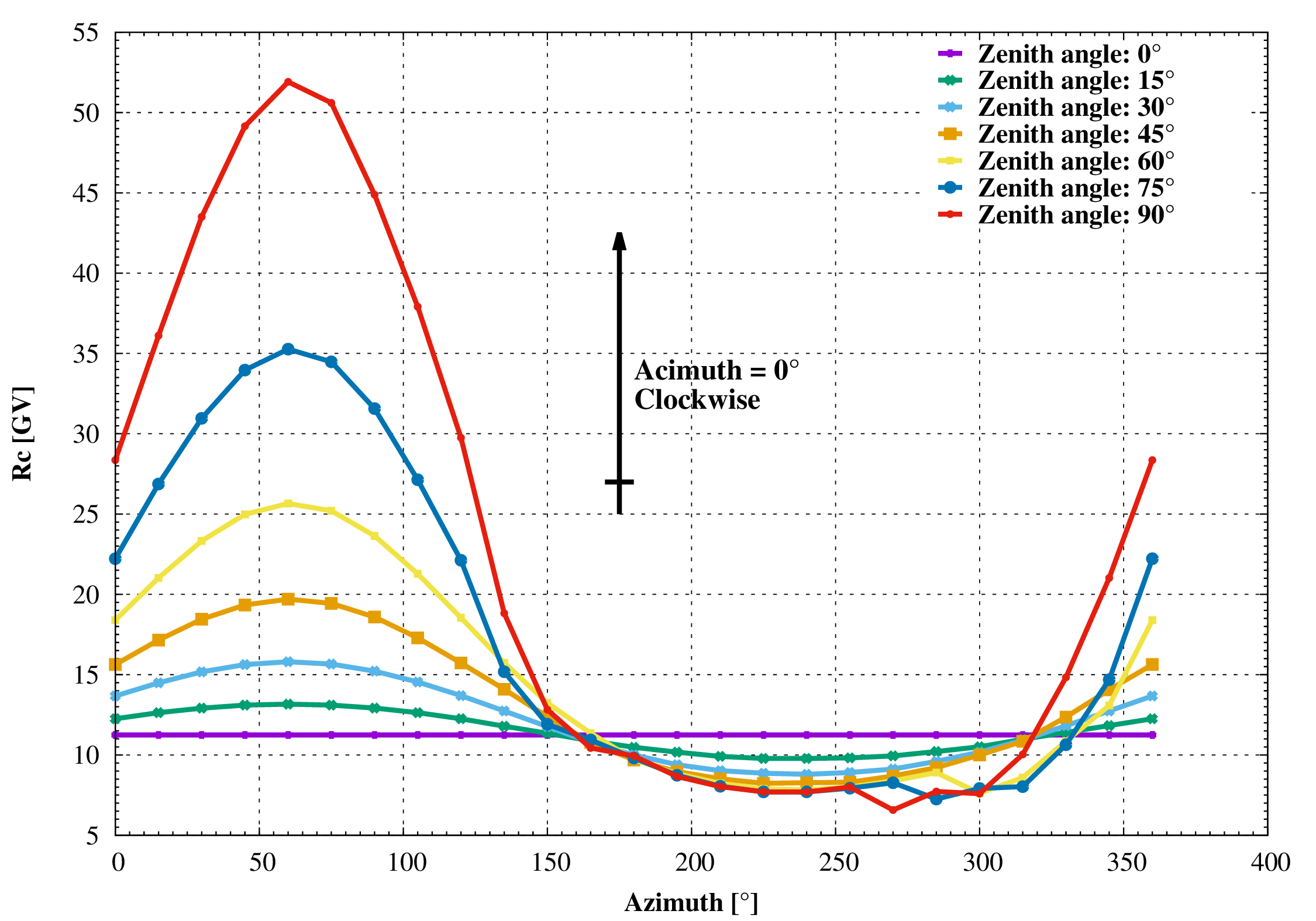}\label{rc20050515109-ce-az} &
\includegraphics[width=0.42\textwidth]{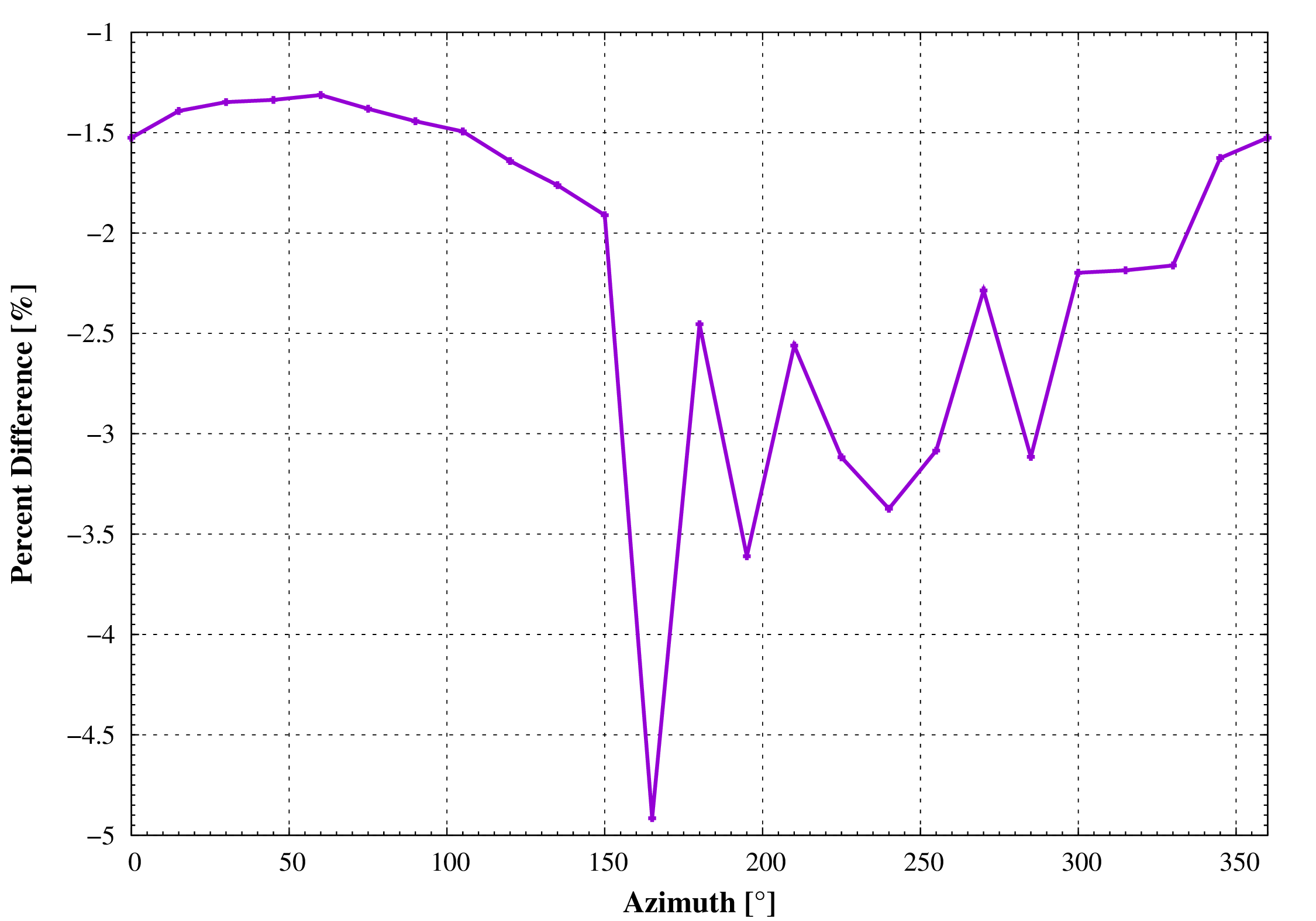}\label{en-diff-rc-secular-dst-247-cenit45} \\
(a) & (b)
\end{tabular}
%%%
  \caption{(a) Directional rigidity cut-off ($R_c$) at the atmosphere edge, as
	  a function of the azimuth angle $\phi$ and for different zenith angles
	  $\theta$ for the LAGO site in Bucamaranga, Colombia, during secular
	  conditions of the geomagnetic field. The effect of the SAA is observed
	  for inclined particles in the range $250^\circ < \phi < 300^\circ$. (b)
	  Relative difference for $\theta=45^\circ$ between the $R_c$ under secular
	  condition (left) and during the geomagnetic storm occurred on May 15th
	  2005 09:00 UTC.  }
  \label{rigidity-results}
\end{figure}

\subsection{Secondary Particles Flux at Ground Level}

The second set of simulation correspond to CORSIKA \footnote{v7.3500 with
QGSJET-II-04; GHEISHA-2002; EGS4; Curve, External Atmosphere, Volumetric
Detector} \citet{HeckEtal1998} of the secondaries observed at ground level by
the EAS produced during the interaction with the atmosphere of the complete
flux of primaries. The flux of all nuclei primaries in the range $1 \leq Z \leq
26$ was assumed to be uniform in solid angle and considering a single power law
for each nuclei to describe the energy dependence in the range $Z \times R_c
\leq E \leq 1$\,PeV \citep{Asorey2011a, DassoAsorey2012, Asorey2013,
AsoreyEtal2015B}. The corresponding parameters were obtained from
\citep{Grieder2001}. As the $R_c$ depends on the arrival direction of each
primary, the total number of primaries $N$ is strongly angular dependent.

\begin{figure}[htb!]
\begin{tabular}{cc}
\includegraphics[width=0.42\textwidth]{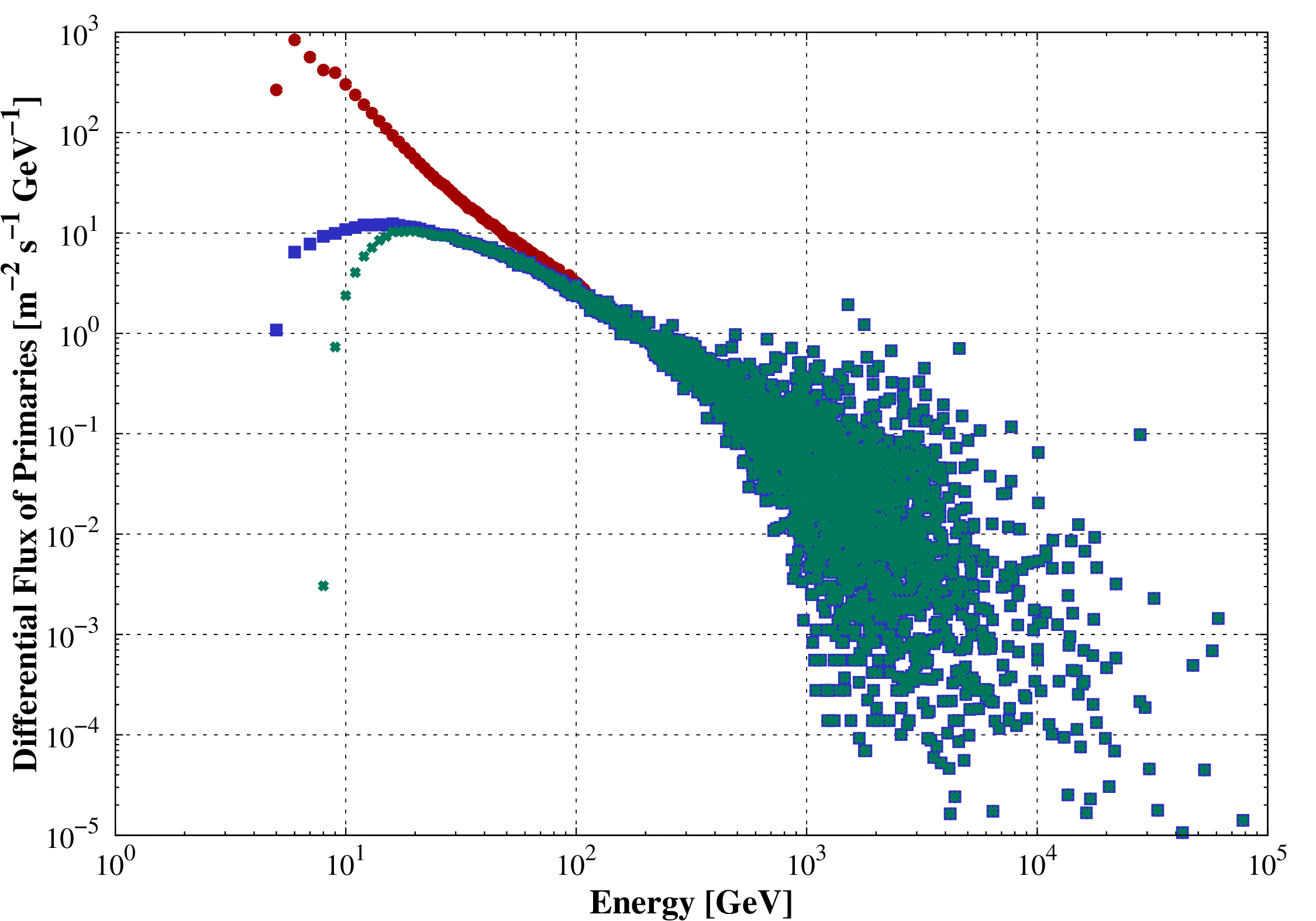}&
\includegraphics[width=0.42\textwidth]{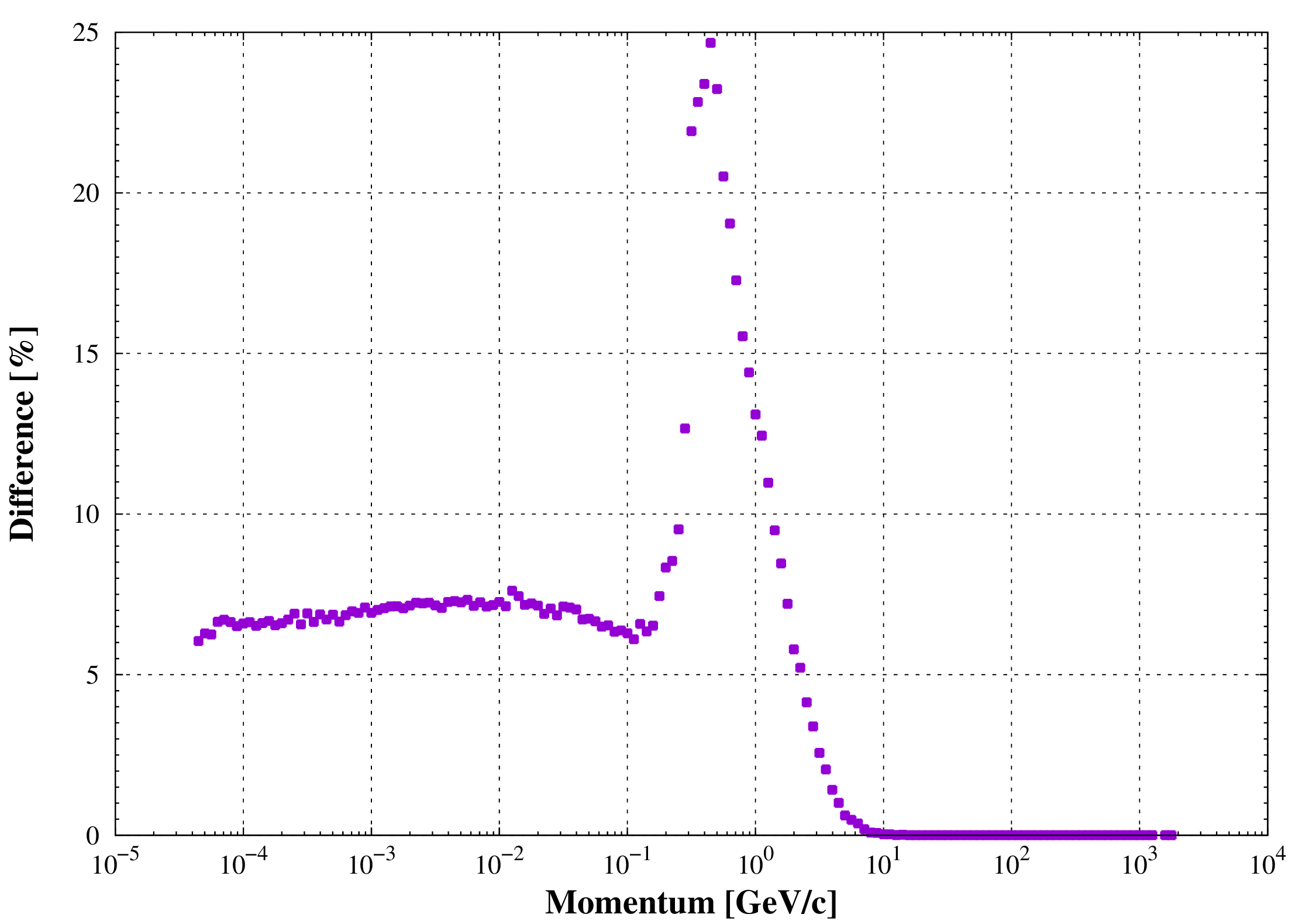} \\
(a) & (b)
\end{tabular}
%%%
  \caption{(a) Geomagnetic effects in the flux of low energy GCR, observed when
	  comparing the flux of primaries that actually produced secondary
	  particles at ground when GF corrections for secular conditions are
	  included (green stars) and when they are not considered (blue squares).
	  The total flux of injected primaries is also included (red circles).  (b)
	  The GF effect in the secondary particles at ground level as a function of
	  the secondary momentum. In this curve, the relative difference between
	  the flux without GF corrections and the corrected flux is shown. The
	  larger differences are observed at $p_{\mathrm{sec}} \simeq 500$\,MeV/c.
	  In this region, the flux of secondaries is dominated by neutrons.
  }
  \label{results-eas}
\end{figure}

The effect of the geomagnetic field in the flux of primaries impinging the
Earth atmosphere can be observed in Fig. \ref{results-eas}(a), where a
comparison of fluxes with/without GF before the edge of the atmosphere is
shown. Only primaries that actually produced secondary particles at ground are
considered. Low energy nuclei are affected, but it is not a sharp cutoff as the
studied effect depends on its $Z$, on the arrival directions of the primaries,
and also includes the effect of atmospheric absorption of secondary particles
during the EAS development.

Figure \ref{results-eas}(b) is a comparison of the expected flux without
corrections with the corresponding geomagnetically corrected one and  relative
difference between these two cases is shown as a function of the particle
momentum. The observed differences are large enough to justify the
incorporation of geomagnetic corrections when low energy secondary flux is
calculated. A large deviation in the total flux is also observed, which is
peaked at $p_{\mathrm{sec}} \simeq 500$\,MeV/c. At this values, the secondary
flux is dominated by neutrons. For this particular component, the decrease at
ground level due to GF effect in secular conditions represents a diminution of
$-36.6\%$. This could be an indication of the sensitivity of secondary neutron
flux as a proxy of the changing conditions in the near-earth space environment.

\subsection{Multi-spectral Analysis of Space Weather Phenomena}

\begin{figure}[htb!]
\begin{tabular}{ccc}
\includegraphics[width=0.31\textwidth]{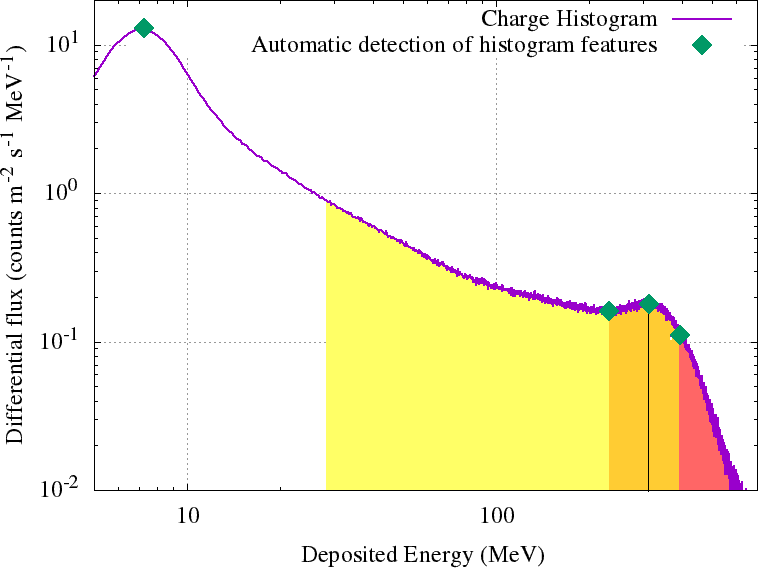}&
\includegraphics[width=0.34\textwidth]{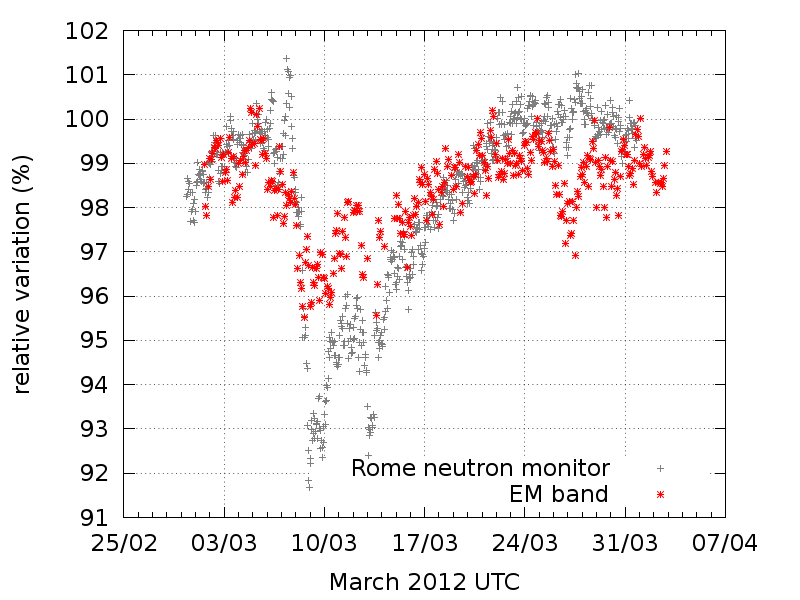}&
\includegraphics[width=0.34\textwidth]{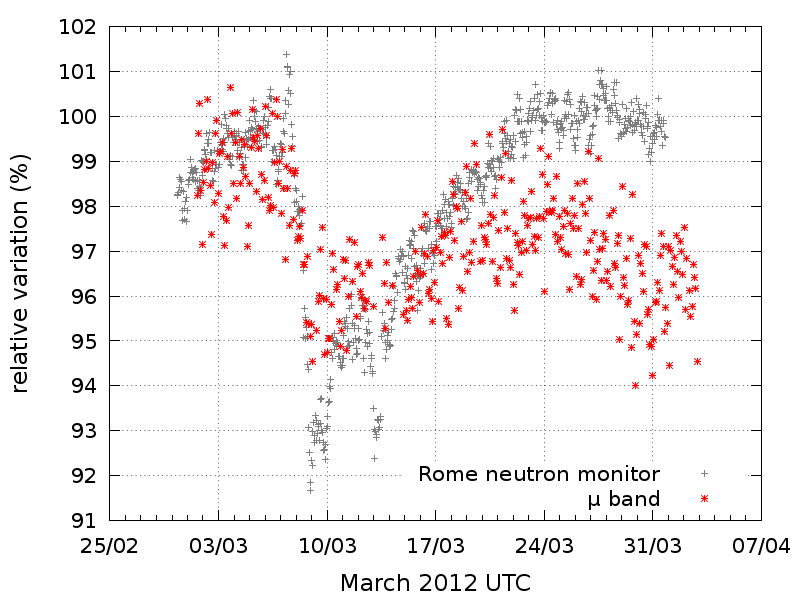} \\
(a) & (b) & (c)
\end{tabular}
  \caption{(a) Charge histogram of one LAGO WCD in Bariloche, Argentina. A fully
	  automated algorithm look for histogram features (green diamonds) to
	  define integrations bands (shaded regions). Each integration band is
	  dominated by different type of particles: EM particles (yellow), muons
	  (orange) and multiple particles (pink). Plates (b) and (c) display results of Multi-Spectal Analysis of the Forbush Decrease of March 8th, 2012 measured in a single $1.8$\,m$^2$ WCD installed in Bariloche,
	  Argentina (red stars), compared with measurements of the Rome neutron
	  monitor (gray pluses): Electromagnetic-band (a), and $\mu$-band (b).
  }
  \label{EM_FD_gen}
\end{figure}

The charge histogram of the WCD, typically used for detector calibration (see
for example \cite{BertouEtal2006}), is obtained by time integration of the
individual pulses measured in the WCD. The resulting histogram, showed in
figure \ref{EM_FD_gen}(a), is originated by the convolution of the response of
the WCD to the different types of EAS particles and the water quality, the
inner coating used, the PMT size and the detector geometry. By doing first
principle calculations and detailed simulations, it is well established that
the small signal region, corresponding to low values of deposited energy in the detector volume, $E_d$, is dominated by
the electromagnetic component (EM, $\gamma$s and $e^\pm$) of the shower; while
high values of $E_d$ correspond to the simultaneous entrance (within the
electronic sampling time, $10$\,ns-$40$\,ns) of multiple particles to the
detector volume (the so called mini-shower - MS - regime). Finally,
intermediate values of $E_d$, evidenced by a characteristic peak
\footnote{The first peak seen at the histogram at very low signals is
	originated by the detector trigger system and is not considered in these
	analysis.
} 
in the histogram called the muon hump, are dominated by single muons through
the detector. By using these histogram features we are able to determine the flux of
secondary particles at different bands of deposited energy in the detector by
using pulse shape discrimination techniques. This is what we called the
multi-spectral analysis technique (MSAT).

We applied the MSAT to analyze the Forbush event associated with the passage of
an interplanetary coronal mass ejection detected on March 8th, 2012, by
different instruments \citep{AlekseenkoEtal2012}. The studied data were
acquired in a single $1.8$\,m$^2$ WCD (with $2.5$\,m$^3$ of pure water and a
$8"$ Hamamatsu R5912 PMT) at the LAGO site of Bariloche, Argentina. After
applied flux pressure correction to each band, the Forbush decrease is clearly
visible in our MSAT, with a maximum peak-to-peak decrease of $\sim 5\%$, $\sim
6\%$, and $\sim 4\%$ in the EM, $\mu$ and MS bands respectively. The temporal
evolution of the EM and $\mu$ bands are shown in Fig. \ref{EM_FD_gen}(b) and
\ref{EM_FD_gen}(c), where, respectively, our data is compared with the one of
Rome neutron monitor\footnote{Available at the NEST repository,
\texttt{http://www.nmdb.eu/nest} }.  Solar daily modulation of the flux is also
visible on both bands. The particular features observed in our data are
currently under careful analysis and will be published soon.

\subsection{Cosmic Rays Induced Background Radiation On Board of Commercial
Flights}

In this subsection we apply the techniques described before to determine the total
integrated flux of cosmic radiation which a commercial aircraft is exposed to
along specific flight trajectories\citep{PinillaAsoreyNunez2015}. To study the
radiation background during a flight and its modulation by effects such as
altitude, latitude, exposure time and transient magnetospheric events, we
perform simulations based on Magnetocosmics and CORSIKA codes. 
%, the former
%designed to calculate the geomagnetic effects on cosmic rays propagation and
%the latter allows us to simulate the development of extended air showers in the
%atmosphere. 

As before, the expected flux of secondary particles in any place along the
plane trajectory is based on the simulation of the complete flux
of cosmic rays primaries within a given range of energy, that includes the effect of the rigidity cut-off at different locations in the Earth, that we summarize here: 
\noindent 
\begin{enumerate}%[leftmargin=*]
  \item Simulation of showers at different altitudes using \textit{CORSIKA}.
    Features of injected primaries at the top of the atmosphere:
    \noindent 
    \begin{itemize}%[leftmargin=*]
      \item Primary nuclei injected: $1\leq Z_p\leq 26$, $1\leq A_p\leq 56$
      \item Very low initial rigidity cut-off rigidity: $R_c=4GV$
      \item Energy and arrival direction: $(R_c\times Z_p)\leq (E_p/GeV)\leq
        10^6$, $0^\circ\leq\theta_p\leq 90^\circ$, $0^\circ\leq\phi_p\leq
        360^\circ$
      \item Simulation time: $t=7200 s$ (primary particles flux is constant and
        isotropic)
    \end{itemize}
  \item Selection and discretization of routes.
  \item Computation of rigidity cut-offs for each point in the trajectory using
    \textit{Magnetocosmics}.
  \item Filter secondary particles by the primary rigidities and the rigidity
    cut-off computed for each point of the trajectory: all those showers
    generated by primary particles with rigidities below the cut-off are simply
    discarded.
  \item Computation of the total amount of particles that hit the aircraft, by
    point-to-point integration of the flux of secondaries along the flight
    trajectory.
\end{enumerate}

\begin{figure}[htb!]
\begin{tabular}{cc}
\includegraphics[width=0.40\textwidth]{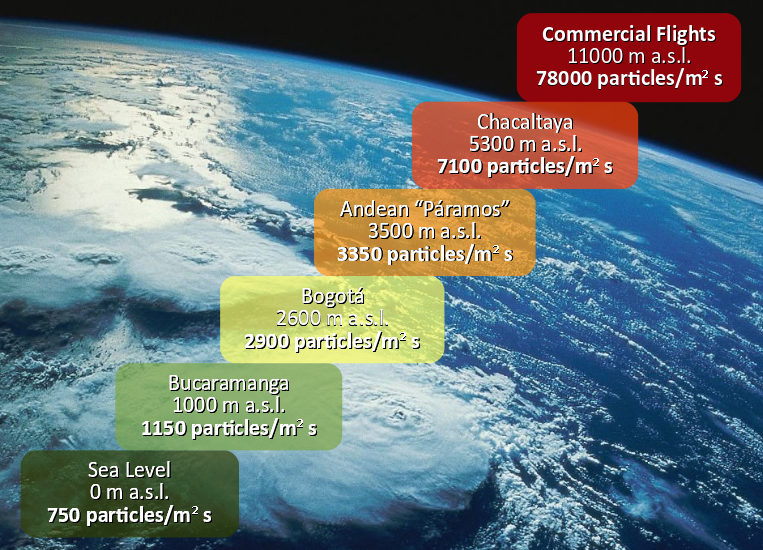} &
\includegraphics[width=0.43\textwidth]{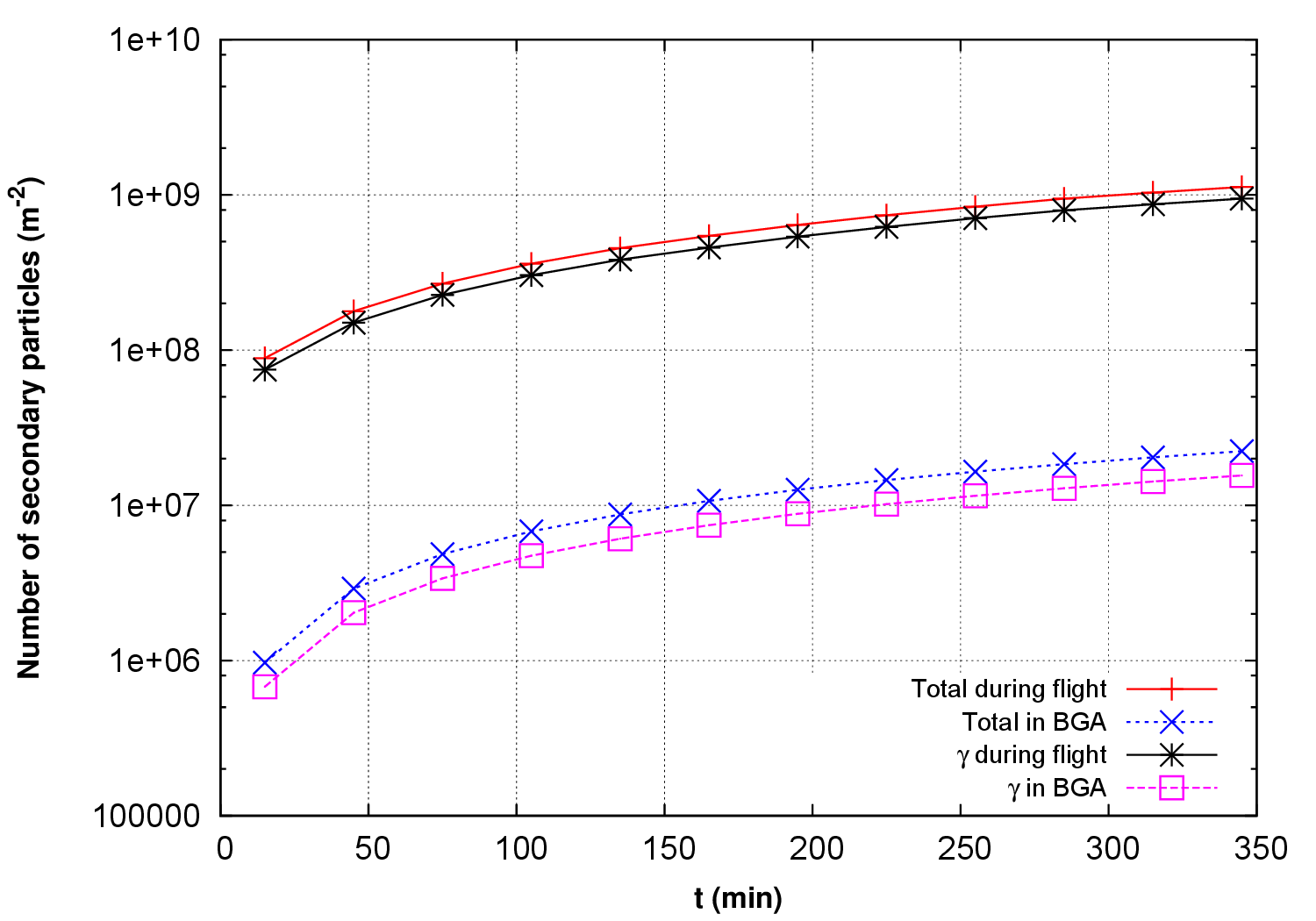} \\
(a) & (b)
\end{tabular}
%%%
  \caption{(a) Total particle flux as function of the altitude at different
	  places on Earth. The atmospheric effect is large enough at high altitudes
	  to justify a detailed simulation at flight level. (b) Integrated flux of
	  particles (red plus signs) and only photons (black asterisks) as a
	  function of time for the flight BOG-EZE, without taken into account the
	  fuselage shield and the effects produced by takeoff and landing. As a
	  comparison, we show the total integrated flux (blue crosses) and photons
	  (magenta squares) for the same calculation but staying the same time at
	  the city of Bucaramanga (Colombia), at an altitude of $1$\,km a.s.l.
	  There is up to two orders of magnitude in the integrated exposure between
	  those two considered cases.
  }
  \label{particleflux}
\end{figure}

As a first estimation, we chose the trajectory of the flight AR1360 BOG-EZE
(Bogot\'a-Buenos Aires), and can be seen elsewhere\footnote{See for example
http://www.flightradar24.com/data/flights/ar1360/}. This route was divided into
$12$ intervals of equal flight time ($\sim 30$ minutes each) and the flux of secondaries along each of
them was assumed to be constant and equal to the flux in the midpoint. The
shield due to the flight fuselage and the effects of takeoff and landing on the
flux was not included in this preliminary analysis (i.e., the aircraft was
supposed to fly at a constant altitude of $11$ km along the whole trajectory).
For each one of this intermediate points, the geomagnetic rigidity cut-off was
calculated by using the method described in the previous section. The result of
this calculation can be seen in figure \ref{particleflux}, where
we show the integrated flux of total particles and photons as a function of
time expected on board of a commercial flight. As a comparison, in the same
figure we show the integrated flux expected for the same calculation but
staying for the same time at the city of Bucaramanga (Colombia) at an altitude
of $1000$\,m a.s.l. In the same figure the effect of the atmospheric absorption
on the EM component is clearly visible as a diminish of the photon flux at
Bucaramanga when compared with the total flux respect to the diminution at
$11$\,km a.s.l. A significant difference in the integrated flux of secondary
particles have been obtained when compared typical flight level altitudes with
ground level. Further investigations are been carried out to include several
other factors, such as the fuselage effect and the conversion to equivalent
doses in living tissues by using, e.g., Geant4 anthropomorphic phantoms (see for
example \cite{matthia2013organ}).

\section{DATA ACCESSIBILITY, REPRODUCIBILITY AND TRUSTWORTHINESS WITH LAGO DATA REPOSITORY}

The geographic distribution of our detectors (see figure
\ref{LAGOSitesChacaltaya}), some of them located at very remote sites,
introduce outstanding challenges for the reliable transference of high volumes
of data. Moreover, this data is acquired on WCD of different geometries and
characteristics. In this section we shall describe the ecosystem of data tools
and services that we had been implemented to help solving the Data
Accessibility, Reproducibility and Trustworthiness (DART) challenge in LAGO
\cite{AsoreyEtAl2015A}. 

The DART initiative was launched by
CHAIN-REDS\footnote{http://www.chain-project.eu} (Coordination and
Harmonisation of Advanced e-infrastructure for Research and Education Data
Sharing): an European Commission co-project focused on promoting and supporting
technological and scientific collaboration across different communities in
various continents\citep{BarberaEtal2014A,BarberaEtal2014B}. This initiative
provided a set of interrelated tools and services, based on worldwide adopted
standards, to provide easy/seamless access datasets, data/documents
repositories and the applications that could generate and/or make use of them.

\textit{Trustworthiness} can be associated to data curation, particularly on
the quality of the metadata describing the experimental protocol and data
provenance\citep{SimmhanPlaleGannon2005}, while \textit{reproducibility} and
\textit{replicability} are closely connected to the accessibility to  data
sources and the possibility to manipulate/analyze data contained in
them\citep{CooperVikWaltemath2015, Peng2011}.

CHAIN-REDS approach to data trustworthiness and reproducibility is based on the
%integration of computational resources and services, with three main
cornerstones:
\noindent 
\begin{enumerate}
	\item adoption of standards for data discoverability, provenance and
		recoverability.%: OAI-PMH\footnote{http://www.openarchives.org/pmh/} for
		%metadata retrieval, Dublin Core\footnote{http://dublincore.org}, as
		%metadata schema, SPARQL\footnote{http://www.w3.org/2001/sw/wiki/SPARQL}
		%for semantic web search and XML\footnote{http://www.w3.org/XML/} as
		%potential standard for the interchange of data;
	
	\item enablement of datasets authorships and user authentication with the
		corresponding assignments of specific roles on data services, which can
		be implemented by two strategies: assignment of Persistent Identifiers
		(PIDs)\citep{Hakala2010} to name data in a unique and timeless manner,
		ensuring that future changes on URIs or internal organization of
		databases will be transparent to the user; and implementation of a
		federated identity provision, a secure, flexible and portable
		mechanism to access e-infrastructures worldwide, based on agreements
		and standards\citep{BarberaEtal2014A}. 
	
	\item access to a plethora of computing power to analyze the retrieved data
		or to contrast them to simulations  through an intuitive web-interface.
		%Over the last years, Science
		%Gateways\citep{AlamedaEtal2007,BarberaFargettaRotondo2011} have risen
		%as an ideal tool to allow scientists across the world to seamlessly
		%access different ICT-based infrastructures for research activities to
		%support their day-by-day work and do better (and faster) research.  
\end{enumerate}

\section{THE PAS PROJECT: ASTROPARTICLE PHYSICS AT THE COLOMBIAN PARAMO}

Motivated by all these developments, and aimed by the international interest to
develop a near equatorial site devoted for the detection of high and ultra high
energy cosmic rays \citep{Watson2014}, we started the development of the PAS
(Social Astronomical Pole, {\emph{Polo de Astronom\'ia Social}})
\citep{asorey2014pas}. It has two main objectives: to build a World class
centre in astroparticle physics and related sciences, and at the same time, to
become a permanent link between Science and Society.

Due to its geographic characteristics, the Andean {\textit{P\'{a}ramo}} located
near Berl\'{\i}n, Colombia ($+7.13$\,N, $-72.9$\,W, $3450$\,m a.s.l.), is an
excellent location to build an array of particle detectors to study cosmic rays
in a wide energy range, including solar activity modulation of cosmic rays,
gamma astronomy, and the high energy region of the cosmic rays spectrum.
Moreover, it can be easily extended to the highest energy region. 

The design of the detector array is based on CORSIKA and GEANT4-based detector
response simulations, and deeply supported by the design experiences of
previous arrays of WCD, such as the Pierre Auger Observatory.  As the number of
computational resources rapidly increases with the primary energy, several
statistical techniques have been developed to simplify those calculations. It
is common to implement the so called thinning algorithms \citep{Billoir2008} to
reduce the number of secondary particles by assigning weights to representative
particles in the evolution of the cascade. However, since this is a compression
method with loss of information, it is required to recover the original flux of
secondary particles without introducing artificial biasses. The so called
{de-thinning} method \citep{StokesEtal2012} is one of the existent methods
designed to deal with this information loss. To validate the first calculations
at the Cosmic Ray knee energy level and beyond, we developed our own
implementation, python based, of the de-thinning method, while introducing some
physical improvements to the original method, such as the inclusion of local
atmospheric models or secondary particle dependant atmospheric interaction
lengths \citep{EstupinanAsoreyNunez2015}. By using this method, we were able to
decrease the computational resources to accomplish the first round of
calculations needed to verify our first principle calculations for the design
of the PAS array.  

\begin{figure}[!!ht]
  \includegraphics[width=0.90\textwidth]{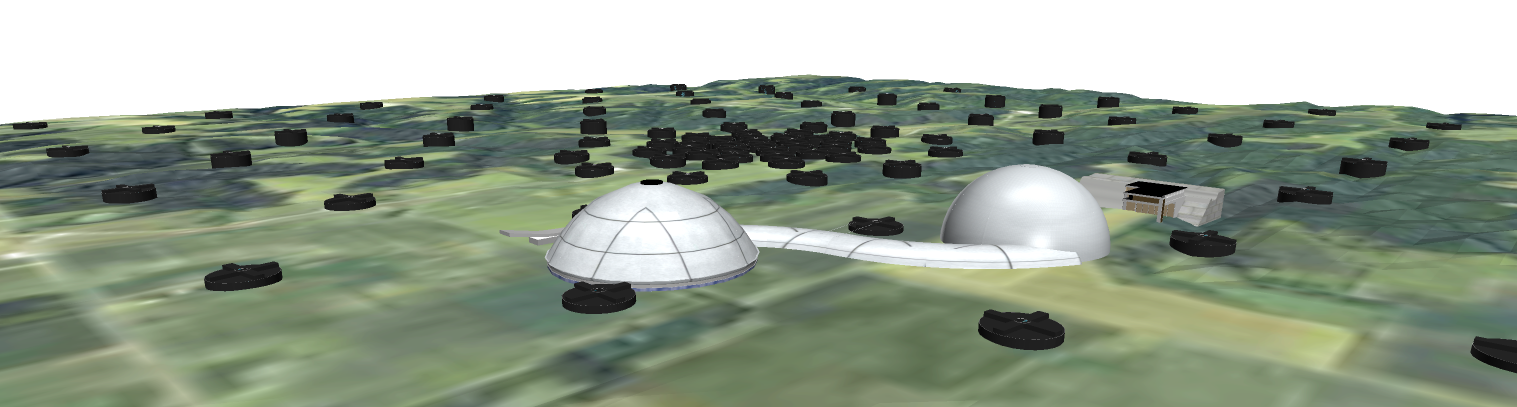}
	\caption{The PAS project concept: the ``Science dome'', the ``Society
dome'' and the ``tunnel of Science'' establishing a bridge, a link, between
Science and Society. An aerial view of the PAS array is also shown: 125 fully
automated and autonomous WCD arranged in a $\sim 16$\,km$^2$ triangular graded
grid.} \label{fig:pas}
\end{figure}

The proposed design of this array, consisting in 125 fully automated and
autonomous WCD arranged in a triangular graded grid and covering a total area
of $\sim 16$\,km$^2$, will allow to implement two different measurement modes:
the counting, LAGO-like, mode and the shower mode. In the counting mode, the
several techniques described before are used to study the flux variation of
particles at detector level. In the shower mode, in contrast, we will look for
time-space correlated signals in different detectors of the array. In this way,
it will be possible to determine the main parameters that characterize the
extensive air shower (EAS) produced by the interaction of a single high-energy
cosmic ray with the atmosphere. The size of the array and the increasing
spacing between detectors will allow to complement present measurements  in the
so called knee region of the cosmic ray energy spectrum ($E \sim
10^{15}$\,eV) and beyond.

Our second objective for PAS will be reached by the installation of three
buildings: one $12$\,m hemispherical dome (``Society''), will harbor a
Convention and Data Visualization Center and a Digital Planetarium; the second
dome (``Science''), will hosts labs and offices for all the scientific
activities, including a $20$\,inches optical fully automatized telescope for
outreach activities and astronomical research at the Paramos. A first sketch of
the design of this facilities is shown in Figure\,\ref{fig:pas}.  The third
building is the one that complete our concept, where an interactive
visualization wall and roof will reproduce different animations and simulations
on different science topics. This is the ``Science tunnel'', which will be the
place where a true link between science and society shall be established.

% Acknowledgement
\section{ACKNOWLEDGMENTS} 

We thank the LAGO Collaboration and Vicerrector\'{\i}a de Investigaci\'on y
Extensi\'on (UIS) for their permanent support. We also thanks to the
Vicerrector\'ia Acad\'emica (UIS) and to ELCIRA (Europe Latin America
Collaborative e-Infrastructure for Research Activities) Project for funding.

\end{document}